\documentclass[10pt]{book}
\def\Bshorttitle{Non-Linear Maximum Entropy Principle }
\def\Bauthor{Tommaso Ruggeri}
\usepackage{hyperref}
\usepackage{amsmath}
\usepackage{amssymb}
\usepackage{amsfonts}
\usepackage{bm}
\usepackage{amsthm}
\usepackage{array,epsfig,fancyheadings}
\usepackage[sectionbib,numbers]{natbib}
\def\BSs #1#2{\pbm[1]{#1}{#1}
\vskip .7cm \centerline{\bf #1}\par \vskip 5pt\noindent{\bf #2}\par\vskip .3cm}

\skip\footins =.6cm
\def\pbm{\pdfbookmark}

\newcolumntype{d}[1]{D{.}{\cdot}{#1}}
\newcolumntype{.}{D{.}{.}{-1}}
\newcolumntype{,}{D{,}{,}{-1}}
\newcommand{\lambdah}{\hat{\lambda}}

\newcommand{\CC}{\mathbf{C}}

\newcommand{\muh}{\hat{\mu}}
\newcommand{\dif}[2]{
\ifx#2\rho{\left(\frac{\partial #1}{\partial \rho}\right)_T}\else{\left(\frac{\partial #1}{\partial T}\right)_\rho}\fi
} 
\newcommand{\pop}{\frac{\Pi}{p}}

\begin{document}
\allowdisplaybreaks \pagestyle{fancy} \thispagestyle{fancyplain}
\lhead[\fancyplain{}\leftmark]{} \chead[]{}
\rhead[]{\fancyplain{}\rightmark} \cfoot{\rm\thepage}
\def\n{\noindent}
\newtheorem{theorem}{Theorem}
\newtheorem{lemma}{Lemma}
\newtheorem{corollary}{Corollary}
\newtheorem{proposition}{Proposition}
\theoremstyle{definition}
\newtheorem{definition}{Definition}
\newtheorem{example}{Example}
\newtheorem{remark}{Remark}
\newcount\fs
\def\dd{\kern 2pt\raise 3pt\hbox{\large .}\kern 2pt\ignorespaces}
\def\fontsizefi{\fontsize{5}{9.5pt plus.8pt minus .6pt}\selectfont}
\def\fontsizesi{\fontsize{6}{10.5pt plus.8pt minus .6pt}\selectfont}
\def\fontsizese{\fontsize{7}{10.5pt plus.8pt minus .6pt}\selectfont}
\def\fontsizeei{\fontsize{8}{12pt plus.8pt minus .6pt}\selectfont}
\def\fontsizeni{\fontsize{9}{13.5pt plus1pt minus .8pt}\selectfont}
\def\fontsizete{\fontsize{10}{14pt plus.8pt minus .6pt}\selectfont}
\def\fontsizeel{\fontsize{10.95}{15pt plus2pt minus .0pt}\selectfont}
\def\fontsizetw{\fontsize{12}{116.5pt plus1pt minus .8pt}\selectfont}
\def\fontsizefo{\fontsize{14.4}{18pt plus1pt minus .8pt}\selectfont}
\def\fontsizesn{\fontsize{17.28}{20pt plus1pt minus 1pt}\selectfont}
\def\fontsizetn{\fontsize{20.73}{26pt plus1pt minus 1pt}\selectfont}
\def\fontsizetf{\fontsize{24.88}{33pt plus1.5pt minus 1pt}\selectfont}
\def\sz#1#2{\fs=#1#2
   \ifnum\fs=05 \fontsizefi
   \else\ifnum\fs=06 \fontsizesi
   \else\ifnum\fs=07 \fontsizese
   \else\ifnum\fs=08 \fontsizeei
   \else\ifnum\fs=09 \fontsizeni
   \else\ifnum\fs=10 \fontsizete
   \else\ifnum\fs=11 \fontsizeel
   \else\ifnum\fs=12 \fontsizetw
   \else\ifnum\fs=14 \fontsizefo
   \else\ifnum\fs=17 \fontsizesn
   \else\ifnum\fs=20 \fontsizetn
   \else\ifnum\fs=25 \fontsizetf
\fi\fi \fi \fi \fi \fi \fi \fi \fi \fi \fi \fi }

\markright{{\scriptsize $\ $}\hfill\break\vskip -16pt {\scriptsize
$\ $}\hfill\break\vskip -16pt {\scriptsize $\ $} \hfill}

\markboth {\hfill{\small \rm \Bauthor}\hfill} {\hfill {\small \rm
\Bshorttitle} \hfill}
\renewcommand{\thefootnote}{}
\abovedisplayskip 12pt
\belowdisplayskip 12pt
\abovedisplayshortskip 5pt
\belowdisplayshortskip 9pt

$\ $\par \vskip .5cm \centerline{\large\bf Non-Linear Maximum Entropy Principle for a} \smallskip
 \centerline{\large\bf   Polyatomic Gas subject to the Dynamic Pressure}\vskip .5cm

\centerline{Tommaso Ruggeri}
 \vskip .5cm
Department of Mathematics and Alma Mater Research Center on Applied Mathematics AM$^2$, University of Bologna, Bologna, Italy  \vskip -4pt
 \noindent
E-mail: \href{mailto:tommaso.ruggeri@unibo.it}{tommaso.ruggeri@unibo.it} \vskip -4pt

\vspace{0.5cm}
\hspace{3cm} \emph{Dedicated to Tai-Ping Liu with esteem and affection}.


\centerline{\bf Abstract}\par \vskip .2cm 

We establish  Extended Thermodynamics (ET) of rarefied polyatomic gases with six independent fields, i.e., 
the mass density, the velocity, the temperature and the dynamic pressure, 
without adopting the near-equilibrium approximation. The closure is accomplished by the Maximum Entropy Principle (MEP) adopting a distribution function that takes into account the internal degrees of freedom of a molecule. The distribution function is not necessarily near equilibrium.   
The result is in perfect agreement with the phenomenological ET theory. To my knowledge, this is the first example of molecular extended thermodynamics with a non-linear closure. The integrability condition of the moments requires that the dynamical pressure should be bounded from below and from above.  In this domain the system is symmetric hyperbolic.
Finally we verify the K-condition for this model and show the existence of global smooth solutions.  \par

\sz11
\setcounter{chapter}{1}                           
\setcounter{equation}{0} 
\BSs{1. Introduction}{} 
Rational extended thermodynamics \cite{MullerRuggeri} (hereafter referred to as ET)  is a thermodynamic theory that is applicable to nonequilibrium phenomena with steep gradients and rapid changes in space-time, which may be out of local equilibrium. It is expressed by a  symmetric hyperbolic system of field equations with the convex entropy.

As ET has been strictly related to the kinetic theory with the closure method of the hierarchy of moment equations, the applicability range of the theory has been restricted within rarefied monatomic gases.  Only recently, however, the ET theory of dense gases and of polyatomic rarefied gases has been successfully developed and has been obtained a $14$-field theory that, in the limit of small relaxation times (parabolic limit), reduces to the Navier-Stokes-Fourier classical theory \cite{ETdense}. This new approach to the case of polyatomic rarefied gases, in particular, is in perfect agreement with the kinetic theory in which the distribution function depends on an extra variable that takes into account the internal degrees of freedom of a molecule \cite{Pavic}. 

Precise modeling of polyatomic gases and of dense gases in nonequilibrium is an active and urgent issue nowadays with many important applications like the study of shock wave structure \cite{ET6shock,Taniguchi-2013,ShockWascom}, which is essentially important, for example, for the atmospheric reentry  problem of a space vehicle.

There are at least three different  methods of closure of moment theory associated with the Boltzmann equation:
\begin{enumerate}
\item A closure at the kinetic level proposed firstly by Grad in the case of $13$ moments, which is based on a perturbative procedure of the distribution function in terms of the Hermite polynomials \cite{Grad};
\item The phenomenological closure of ET by using the universal principles of physics to select admissible constitutive equations \cite{MullerRuggeri,LiuMull,LMR} ; 
\item The kinetic closure of molecular ET  by using  the Maximum Entropy Principle  (MEP) \cite{Dreyer,ET}.
\end{enumerate}
It is very suitable and in some sense surprising that the three different closure methods give the same result in the case of the 13-moment theory for monatomic gases \cite{MullerRuggeri} and also in the $14$-moment theory for rarefid polyatomic gases \cite{Pavic}, provided that the thermodynamic processes are not far from equilibrium.

We here want to focus mainly on MEP and therefore we want firstly to summarize the principle and its limitation.
The principle of maximum entropy 
has its root in statistical mechanics and is developed by
E. T. Jaynes in the context of the theory of information related to the concept of the Shannon entropy   
\cite{Jaynes-1957, Kapur-1989}. 
MEP states that the probability distribution that represents the current state of knowledge in the best way is the one with the largest entropy.
Concerning the applicability of MEP in nonequilibrium thermodynamics, this was
originally motivated by the similarity 
between the field equations in ET and the moment equations, and later  by the 
observation made by Kogan \cite{Kogan-1967} that  
Grad's distribution 
function maximizes the entropy. The MEP was proposed in ET for the first time by Dreyer   \cite{Dreyer}. 
The MEP procedure was 
then generalized by M\"uller  and Ruggeri   to the case of any number of moments \cite{ET}, 
and later proposed again and popularized by Levermore    \cite{Levermore-1996}. In the case of moments associtaed to the Boltzmann equation the complete 
equivalence between the entropy principle and the MEP was finally proved by Boillat   and Ruggeri \cite{Boillat-1997}.  
Later MEP was formulated also in a quantum-mechanical context \cite{Degond-2003,Trovato}. 

As seen below, the truncated distribution function of MEP has the meaning also far from equilibrium provided that the integrals of the moments are convergent.
The problem of the convergence of the moments is one of the main questions in the far-from-equilibrium case. In particular,  the index of truncation of the moments $N$ must be even. This implies that the Grad  theory with $13$ moments is not allowed in a situation far from equilibrium!
For this reason the truncated distribution function is formally expanded in the neighborhood of equilibrium  as a perturbation of the Maxwellian distribution.  
All closures by the MEP procedure are valid only near equilibrium. As a consequence, hyperbolicity exists only in some small domain of configuration space near equilibrium \cite{MullerRuggeri,Brini}.

The aim of this paper is to prove that, in the case of rarefied polyatomic gases, a theory can be established with the closure that is valid even far from equilibrium. This is a theory with 6 independent fields, i.e., 
the mass density, the velocity, the temperature and the dynamic pressure. We will show that this non-linear closure matches completely the  previous result obtained by using only the macroscopic method  \cite{NonLinear6}.

\setcounter{chapter}{2}                           
\setcounter{equation}{0} 
\BSs{2. Rarefied polyatomic Gas}{}

A crucial step in the development 
of the kinetic theory of rarefied polyatomic gases was made by Borgnakke and Larsen \cite{Borgnakke-1975}. 
It is assumed that the distribution function depends on, in addition to the velocity of particles $\mathbf{c}$, 
a continuous variable $I$
representing the energy of the internal modes of a molecule. 
This model was initially used 
for Monte Carlo simulations of polyatomic gases, and later it has been applied to the derivation 
of the generalized Boltzmann equation by Bourgat,     Desvillettes,     Le Tallec     and Perthame     \cite{Bourgat-1994}.
The 
distribution function $f(t,\mathbf{x},\mathbf{c},I)$ is
defined on the extended domain $[0,\infty) \times  {R}^{3}
\times  {R}^{3} \times [0,\infty)$. Its rate of change is
determined by the Boltzmann equation which has the same form as in the case of monatomic gases: 
\begin{equation}
\partial_t f + c_i \, \partial_i f = Q,  \label{pat}
\end{equation}
where the right-hand side, the collision term, describes the effect of collisions between molecules. 
  The collision term $Q(f)$ now takes into account
the existence of the internal degrees of freedom through the collisional cross section. Here $\partial_t= {\partial}/{\partial t}$ and $\partial_i= {\partial}/{\partial x_i}$. 

The idea, firstly proposed at the macroscopic level by Arima, Taniguchi, Ruggeri and Sugiyama \cite{ETdense} 
and successively in the kinetic framework in \cite{Pavic} and \cite{Annals},  is to consider, instead of the typical single hierarchy of moments, a double hierarchy, i.e., 
an  $F$-series at the index of truncation $N$ and a $G$-series at the index $M$: $(N,M)$ system given by
\begin{align*}
\begin{split}
 &\partial_t F + \partial_i F_i = 0,\\
 &\partial_t F_{k_1} + \partial_i F_{ik_1} = 0, \\
 & \partial_t F_{k_1 k_2} + \partial_i F_{ik_1 k_2} = P_{k_1 k_2}, \ \ \quad \quad \qquad \partial_t G_{ll} + \partial_i G_{ill}=0,\\
 &\vdots \hspace{5.05cm}\quad  \qquad \partial_t G_{llj_1} + \partial_i G_{llij_1} =Q_{llj_1},\\
 &\vdots \hspace{5.29cm} \quad \qquad \vdots \\
& \partial_t F_{k_1 k_2 \dots k_N} + \partial_i F_{ik_1 k_2 \dots k_{N}} = P_{k_1 k_2 \dots k_N}, \,\, \vdots \\
&  \hspace{5.2cm} \quad \qquad  \partial_t G_{ll j_1 j_2 \dots j_M} + \partial_i G_{llij_1 j_2 \dots j_{M}} = Q_{ll j_1 j_2 \dots j_M}. 
\end{split}
\end{align*}
with
\begin{equation}
F_{k_1 k_2 \cdots k_p}= \int_{R^{3}} \int_{0}^{\infty}{m f(t, \mathbf{x}, \mathbf{c}, I)  c_{k_1} c_{k_2} \cdots c_{k_p}  \varphi(I) \, dI \,
    d\boldsymbol{c}},  \label{prima}
\end{equation}\begin{equation}
G_{llk_1 k_2 \cdots k_q}= \int_{R^{3}} \int_{0}^{\infty}{m f(t, \mathbf{x}, \mathbf{c}, I) \left( c^2+2\frac{I}{m}\right)  c_{k_1} c_{k_2} \cdots c_{k_q}  \varphi(I) \, dI \,
    d\boldsymbol{c}},  \label{seconda}
\end{equation}
and $0\leq p\leq N, \ 0\leq q\leq M$ (when the index $p=0$ we have $F$ and when $q=0$, $G_{ll}$).
The double hierarchy is composed of the traditional \emph{velocity-moments} $F$'s  and the \emph{energy-moments} $G$'s  where the variable $I$ of the internal modes plays a role. The connection between the index $M$ and $N$ is discussed in \cite{Annals}.
The non-negative measure $\varphi(I) \, dI$ is introduced so as to recover the 
classical caloric equation of state for
polyatomic gases in equilibrium. The functional form of $\varphi$ will be given in the next section.

\sz11
\setcounter{chapter}{2}                           
\BSs{}{2.1 Equilibrium Distribution Function and the Euler System}

\label{sec:Polyatomic-Euler}
Let us consider firstly the case of $5$ moments corresponding to an Euler fluid. In this case, $N=1$ and $M=0$.
The collision invariants in this model form a 5-vector:
\begin{equation}
    m \left(1, c_{i},  c^2 
    +2 \frac{I}{m} \right)^{T},
  \label{Poly:Invariants}
\end{equation}
which leads to hydrodynamic variables:
\begin{equation}
  \left(
  \begin{array}{c}
  F \\
  F_i \\
  G_{ll} \\
   \end{array}%
    \right) =
  \left(%
    \begin{array}{c}  
    \rho \\
    \rho v_{i} \\
    \rho v^2  + 2 \rho \varepsilon \\
    \end{array}
    \right)
  = \int_{R^{3}} \int_{0}^{\infty}
      m \left(%
        \begin{array}{c}
        1 \\
        c_{i} \\
        c^2 + 2 I/m \\
        \end{array}%
      \right)
      f(t, \mathbf{x}, \mathbf{c}, I) \varphi(I) \, dI \,
      d\mathbf{c}, 
  \label{Poly:HydroMoments}
\end{equation}
The symbols are the usual ones: $\rho,v_i,\varepsilon$ are, respectively, the mass density, the i-th component of the velocity and the specific internal energy.
 The entropy is defined by the
relation:
\begin{equation}
  h = - k_B \int_{R^{3}} \int_{0}^{\infty}
    f \log f \varphi(I) \, dI \, d\mathbf{c}.
  \label{Poly:Entropy}
\end{equation}
By introducing the peculiar velocity:
 \begin{equation}
 C_i=c_i-v_i, \label{peculiarv}
 \end{equation}
we rewrite Eq. (\ref{Poly:HydroMoments}) as follows:
\begin{equation}
  \left(%
  \begin{array}{c}
    \rho \\
    0_{i} \\
    2 \rho \varepsilon \\
  \end{array}%
  \right)
  = \int_{R^{3}} \int_{0}^{\infty}
    m \left(%
      \begin{array}{c}
      1 \\
      C_{i} \\
      C^2 + 2 I/m \\
      \end{array}%
    \right)
    f(t, \mathbf{x}, \mathbf{C}, I) \varphi(I) \, dI \,
    d\mathbf{C}.
  \label{Poly:HydroPeculiar}
\end{equation}
Note that the internal energy density can be divided into the
translational part $\rho \varepsilon_{T}$ and the part of the
internal degrees of freedom $\rho \varepsilon_{I}$:
\begin{align}
  \rho \varepsilon_{T} & = \int_{R^{3}} \int_{0}^{\infty}
    \frac{1}{2} m C^2
    f(t, \mathbf{x}, \mathbf{C}, I) \varphi(I) \, dI \,
    d\mathbf{C},
  \nonumber \\
  \rho \varepsilon_{I} & = \int_{R^{3}} \int_{0}^{\infty}
    I f(t, \mathbf{x}, \mathbf{C}, I) \varphi(I) \, dI \,
    d\mathbf{C}.
  \label{Poly:EnergyDef}
\end{align}
The energy $\rho \varepsilon_{T}$ is related to the kinetic temperature $T$:
\begin{equation}
  \varepsilon_{T} = \frac{3}{2} \frac{k_B}{m} T,
  \label{Poly:Energy_T}
\end{equation}
where $k_B$ and $m$ are the Boltzmann constant and the atomic mass, respectively.
The weighting function $\varphi(I)$ is determined in 
such a way that it recovers the caloric equation of state for
polyatomic gases. If $D$ is the degrees of freedom of a
molecule, it can be shown that the relation $\varphi(I) = I^{\alpha}$ leads to the 
appropriate caloric equation of state:
 \begin{equation}
   \varepsilon  =  \frac{D}{2}  
     \frac{k_B}{m} T,
 \qquad   \alpha = \frac{D - 5}{2}.
  \label{Poly:AlphaDj}
\end{equation}

The maximum entropy principle is expressed in terms of the
following variational problem: determine the distribution
function $f(t,\mathbf{x},\mathbf{C},I)$ such that $h \to
\textrm{max}$, under the constraints
(\ref{Poly:HydroMoments}), or equivalently, due to the Galilean
invariance, under the constraints (\ref{Poly:HydroPeculiar}). The
result due to Pavic, Ruggeri and Simi\'c \cite{Pavic} is summarized as follows:
\begin{theorem}
The  distribution function that maximizes the entropy
(\ref{Poly:Entropy}) under the constraints
(\ref{Poly:HydroPeculiar}) has the form:
\begin{equation}
  f_{E} = \frac{\rho}{m \, (k_B T)^{1 + \alpha} \Gamma(1 + \alpha) } \left( \frac{m}{2 \pi k_B T} \right)^{3/2}
    \exp \left\{ - \frac{1}{k_B T} \left( \frac{1}{2} m
    C^2 + I \right) \right\}.
  \label{Poly:Maxwellian}
\end{equation}
\end{theorem}
This is the generalized Maxwell distribution function for polyatomic gases.
In \cite{Pavic}, the following theorem was also proved: 
\begin{theorem}
If (\ref{Poly:Maxwellian}) is the local equilibrium distribution function with
$\rho \equiv \rho(t,\mathbf{x})$, $\mathbf{v}   \equiv
\mathbf{v}  (t,\mathbf{x})$ and $T \equiv  T(t,\mathbf{x})$, then
the hydrodynamic variables $\rho$, $\mathbf{v}  $ and $T$ satisfy
the Euler system:
\begin{eqnarray} \label{Poly:EulerSystem}
&&\frac{\partial \rho }{\partial t} +\frac{\partial }{\partial x_{i}}\left(
\rho v_{i}\right) =0,\notag \\
&&\frac{\partial }{\partial t}\left( \rho v_{j}\right) +\frac{\partial }{%
\partial x_{i}}\left( \rho v_{i}v_{j}+p\delta _{ij}\right) =0, \\
&&\frac{\partial }{\partial t}\left( \rho \varepsilon+\rho \frac{v^{2}}{2}\right) +%
\frac{\partial }{\partial x_{i}}\left\{ \left( \rho\varepsilon+\rho \frac{v^{2}}{2}%
+p\right) v_{i}\right\} =0 \notag
\end{eqnarray}
with 
\begin{equation}
\label{thermal}
p = \frac{k_B}{m} \rho\,  T, \qquad \varepsilon =\frac{D}{2} \frac{k_B}{m}\, T.
\end{equation}
\end{theorem}
This is an important result because we can obtain the Euler equations from the kinetic equation for any kind of polyatomic gases as well as monatomic gases.

\sz11
\setcounter{chapter}{3}                           
\setcounter{equation}{0} 
\BSs{3. The 6 Moment-Equations for Polyatomic Gases}{}

\label{sec:Polyatomic-Hierarchy}
The 14-field theory, $N=2$ and $M=1$, gives us a complete phenomenological model but its differential system is rather complex and the closure is in any way limited to a theory near equilibrium.
Let us consider now 
a simplified theory with 6 fields
(referred to as the ET6 theory): the mass density $\rho$, 
the velocity $\mathbf{v}$, the temperature $T$, and the dynamic (nonequilibrium) pressure $\Pi$.  
This simplified theory preserves the main physical properties of the more complex theory of 14 variables, in particular, when the bulk viscosity plays more important role than the shear viscosity and the heat conductivity do.  
This situation is observed in many gases such as rarefied hydrogen gases and carbon dioxide gases at some temperature ranges \cite{CC,Arima-2012b,Taniguchi-2013}.
ET6 has another advantage to  offer us a more affordable hyperbolic partial differential system. In fact, it is the simplest system that takes into account a dissipation mechanism after the Euler system of perfect fluids.  In the present case we have:
\begin{align}
\begin{split}
 &\frac{\partial F}{\partial t}+\frac{\partial F_k}{\partial x_k}=0,\\
 &\frac{\partial F_i}{\partial t}+\frac{\partial F_{ik}}{\partial x_k}=0, \\
 &{\frac{\partial F_{ll}}{\partial t}+\frac{\partial F_{llk}}{\partial x_k}=P_{ll},} \ \ \ \ \ \ \ \ \ \ \frac{\partial G_{ll}}{\partial t}+\frac{\partial G_{llk}}{\partial x_k}=0\\
\end{split}\label{balanceFGEUl}
\end{align}
with
\begin{equation}\label{FEulero}
  \left(%
    \begin{array}{c}
    F \\
   F_{i} \\
   F_{ll}\\
    \end{array}%
    \right) =
  \left(%
  \begin{array}{c}
  \rho \\
  \rho v_{i} \\
  \rho v^2 + 3(p+\Pi) \\
  \end{array}%
  \right)=
  \int_{R^{3}} \int_{0}^{\infty}
        m \left(%
          \begin{array}{c}
          1 \\
          c_{i} \\
          c^2 \\
          \end{array}%
        \right)
        f I^\alpha \, dI \,
        d\mathbf{c}
\end{equation}
and
\begin{equation}\label{GEulero}
G_{ll}=\rho v^2 + 2 \rho \varepsilon= \int_{R^{3}} \int_{0}^{\infty} m(c^2+2 {I}/{m})f I^\alpha \, dI \,
        d\mathbf{c}.
\end{equation}

\sz11
\setcounter{chapter}{3}                           
\BSs{}{3.1 Nonequilibrium distribution function}
We want to prove the following theorem:
\begin{theorem}
The  distribution function that maximizes the entropy (\ref{Poly:Entropy})
 under the constraints (\ref{FEulero})
(\ref{GEulero}) has the form:
\begin{eqnarray}
 f  &=& \frac{\rho}{m \, (k_B T)^{1 + \alpha} \Gamma(1 + \alpha) } \left( \frac{m}{2 \pi k_B T} \,\frac{1}{1+\frac{\Pi}{p}}  \right)^{3/2}
  \left(\frac{1}{1-\frac{3}{2( 1+\alpha)}  \frac{\Pi}{p}}\right)^{1+\alpha}\nonumber \\
&&    \exp \left\{ - \frac{1}{k_B T} \left( \frac{1}{2} m
    C^2\left(\frac{1}{1+\frac{\Pi}{p}}\right) + I \left(\frac{1}{1-\frac{3}{2(
 1+\alpha)}
    \frac{\Pi}{p}}\right)\right) \right\}.
     \label{f6}
    \end{eqnarray}
    All the moments are convergent provided that 
\begin{equation}
    -1<\frac{\Pi}{p} < \frac{2}{3}(1+\alpha), \qquad \alpha>-1.
    \end{equation}
 \end{theorem}
\textbf{Proof}:
The proof of the theorem is accomplished with the use of the Lagrange
multiplier method. Introducing the vector of the 
multipliers $ \left( \lambda , \lambda_{i} ,\lambda_{ll} , \mu_{ll}
\right)$, we define the functional:
\begin{eqnarray*}
&&  \mathcal{L} =- \int_{R^{3}} \int_{0}^{\infty} k_B f \log f\,  I^\alpha\, dI \,
    d\mathbf{c} +\lambda  \left(\rho -  \int_{R^{3}} \int_{0}^{\infty} m  f \, I^\alpha\, dI \,d\mathbf{c} \right)+ \\
& &   + \lambda_i  \left(\rho v_i - \int_{R^{3}} \int_{0}^{\infty} m f c_i\,  I^\alpha\, dI \, d\mathbf{c} \right)+   
 \lambda_{ll}  \left(\rho v^2 +3(p+\Pi) -  \int_{R^{3}} \int_{0}^{\infty}m c^2 f\,  I^\alpha\, dI \, d\mathbf{c} \right) + \\
&& + \mu_{ll}  \left(\rho v^2 +2 \rho \varepsilon -  \int_{R^{3}} \int_{0}^{\infty}m \left(c^2+2 \frac{I}{m}\right) f\,  I^\alpha\, dI \, d\mathbf{c} \right).
\end{eqnarray*}
As this is a functional of the distribution function $f$ and we want to maximize it with respect to $f$ with the given macroscopic quantities, this functional can be substituted by the following one: 
\begin{eqnarray}
&&  \mathcal{L} =- \int_{R^{3}} \int_{0}^{\infty} k_B f \log f\,  I^\alpha\, dI \,
    d\mathbf{c} -\lambda    \int_{R^{3}} \int_{0}^{\infty} m  f \, I^\alpha\, dI \,d\mathbf{c} -\nonumber \\
& &  -  \lambda_i   \int_{R^{3}} \int_{0}^{\infty} m f c_i\,  I^\alpha\, dI \, d\mathbf{c} -   
 \lambda_{ll}    \int_{R^{3}} \int_{0}^{\infty}m c^2 f\,  I^\alpha\, dI \, d\mathbf{c} - \label{L6v}\\
&& - \mu_{ll}    \int_{R^{3}} \int_{0}^{\infty}m \left(c^2+2 \frac{I}{m}\right) f\,  I^\alpha\, dI \, d\mathbf{c}. \nonumber
\end{eqnarray}
Since $\mathcal{L}$ is a scalar, it must retain the same value
in the case of zero hydrodynamic velocity $\mathbf{v}   = \mathbf{0}$ due
to the Galilean invariance. Therefore:
\begin{eqnarray}
&&  \mathcal{L} =- \int_{R^{3}} \int_{0}^{\infty} k_B f \log f\,  I^\alpha\, dI \,
    d\mathbf{C} -\hat{\lambda}    \int_{R^{3}} \int_{0}^{\infty} m  f \, I^\alpha\, dI \,d\mathbf{C} - \nonumber \\
& &   \hat{ \lambda}_i   \int_{R^{3}} \int_{0}^{\infty} m f C_i\,  I^\alpha\, dI \, d\mathbf{C} -   
 \hat{\lambda}_{ll}    \int_{R^{3}} \int_{0}^{\infty}m C^2 f\,  I^\alpha\, dI \, d\mathbf{C} - \label{L6sv}\\
&&\hat{\mu}_{ll}    \int_{R^{3}} \int_{0}^{\infty}m \left(C^2+2 \frac{I}{m}\right) f\,  I^\alpha\, dI \, d\mathbf{C}.\nonumber
\end{eqnarray}
 Comparison between (\ref{L6v}) and
 (\ref{L6sv}) 
  yields the relations between the Lagrange multipliers and the corresponding zero-velocity Lagrange multipliers indicated by hat:
 \begin{equation}\label{lambdalino}
   \lambda  = \hat{\lambda} - \hat{\lambda}_{i}v_i
     +  (\hat{\lambda}_{ll} +\hat{\mu}_{ll} ) v^2 ;
     \quad
   \lambda_i  = \hat{\lambda}_{i}
     - 2  (\hat{\lambda}_{ll} +\hat{\mu}_{ll} ) v_i;
     \quad \lambda_{ll}= \hat{\lambda}_{ll} \quad
   \mu_{ll} = \hat{\mu}_{ll},
 \end{equation}
 which dictate the velocity dependence of the Lagrange multipliers. We notice that these relations are in
 accordance with the general results of the Galilean invariance \cite{Galileo}.
The Euler-Lagrange equation $\delta \mathcal{L}/\delta f = 0$
leads to the following form of the distribution function:
\begin{equation}
 f = \exp^{-1 - \frac{m}{k_B}\chi}, \label{gff6}
\end{equation}
where
\begin{equation*}
 \chi = \hat{\lambda} + \hat{\lambda}_i C_i+\hat{\lambda}_{ll}C^2 + \hat{\mu}_{ll} \left(C^2 + 2\frac{I}{m}\right).
\end{equation*}
By introducing the following variables:
\begin{equation}
 \xi = \frac{m}{k_B}(\lambdah_{ll} + \muh_{ll}) ,\quad \eta_i= \frac{m}{k_B} \hat{\lambda}_i, \quad
 \zeta =   \frac{2}{k_B}\muh_{ll} ,\quad
 \Omega =  \exp \left(-1-\frac{m}{k_B}\lambdah\right),  
 \label{defval}
\end{equation}
the distribution function can be rewritten as 
\begin{align}
 f = \Omega  \mathrm{e}^{-\zeta I }\mathrm{e}^{- \xi C^2-\eta_i C_i }. \label{dis}
\end{align}
Inserting (\ref{dis}) into the second equation of (\ref{FEulero}) evaluated at the zero velocity, we obtain immediately $\eta_i=0$.  Then the remaining equations of (\ref{FEulero}) and (\ref{GEulero}) evaluated for $\mathbf{v}=0$ become
\begin{eqnarray}
 &&\rho  = \int_{R^3} \int_0^{\infty} m	f  I^\alpha \, dI \, d\CC  
  = m \pi ^{3/2} \Gamma (1+\alpha) \frac{\Omega }{\xi ^{3/2} \zeta^{1+\alpha}}, \nonumber\\
 && p +\Pi  = \frac{1}{3}\int_{R^3} \int_0^{\infty} m f C^2 I^\alpha \, dI \, d\CC 
 = m \pi ^{3/2} \Gamma(1+\alpha)\frac{\Omega}{2 \xi ^{5/2} \zeta^{1+\alpha}},\label{densities}\\
 & &\rho \varepsilon  = \int_{R^3} \int_0^{\infty} m f \left(\frac{C^2}{2} + \frac{I}{m}\right) I^\alpha \, dI \, d\CC = \nonumber\\
 & & = m \pi ^{3/2} \Gamma(1+\alpha) \frac{\Omega  }{4 \xi^{5/2}\zeta^{1+\alpha}}\left(3+ \frac{4}{m} (1+\alpha) \frac{\xi}{\zeta}\right).  \nonumber
\end{eqnarray}
From the integrability condition, we have  
\begin{equation}\label{converg}
\zeta>0, \qquad \xi >0, \qquad \alpha>-1.
\end{equation}
From  (\ref{densities}) and (\ref{Poly:AlphaDj}), we obtain
\begin{align}
 \begin{split}
 &\varepsilon = \frac{1}{4\xi} \left\{3+\frac{2}{m}(D-3)\frac{\xi}{\zeta}\right\},\\
 &p = \frac{m}{2D}\pi^{3/2} \Gamma\left(\frac{D-3}{2}\right)\frac{\Omega}{\xi^{5/2}\zeta^{\frac{D-3}{2}}} \left\{3+\frac{2}{m}(D-3)\frac{\xi}{\zeta}\right\},\\
 &\Pi =  \frac{m}{2}\pi^{3/2} \Gamma\left(\frac{D-3}{2}\right) \frac{D-3}{D}\frac{1 - \frac{2}{m}\frac{\xi}{\zeta}}{\xi^{5/2}\zeta^{\frac{D-3}{2}}}\Omega.  
 \end{split}
\label{r2}
\end{align}
We can invert these relations as follows:
\begin{align}\label{xizeta}
 \begin{split}
 &\xi = \frac{\rho }{2 p} \frac{1}{1+\pop},\\
 &\zeta =  \frac{\rho}{m}\,\,\frac{(D-3)}{2\rho \varepsilon -3(p+\Pi)}  = \frac{\rho}{mp}\,\frac{1}{1-\frac{3}{D-3}\pop},\\
 & \Omega = \frac{\rho}{m \pi ^{3/2} \Gamma\left(\frac{D-3}{2}\right)}
 \left(\frac{\rho}{2p}\frac{1}{1+\pop }\right)^{\frac{3}{2}}\left(\frac{\rho}{mp}\frac{1}{1 - \frac{3}{D-3}\pop}\right)^{\frac{D-3}{2}}.
 \end{split}
\end{align}

The integrability conditions (\ref{converg}) imply that, for a bounded solution, the ratio $\Pi/p$ must satisfy 
\begin{equation}
 p+\Pi >0 \quad \text{and} \quad (D-3)p - 3\Pi >0.
\end{equation}

Inserting (\ref{xizeta}) into the distribution function (\ref{dis}), we obtain (\ref{f6}) and the proof is completed. When $\Pi \rightarrow 0$ the (\ref{f6})  becomes the equilibrium distribution function (\ref{Poly:Maxwellian}).

\sz11
\setcounter{chapter}{3}                           
\BSs{}{3.2 Closure and Field equations}
Substituting (\ref{f6}) into the fluxes $F_{llk}, G_{llk}$ and into the production term $P_{ll}$ of (\ref{balanceFGEUl}), we obtain after some calculations
\begin{align}\label{chiuso}
\begin{split}
& F_{ik} =\int_{R^{3}} \int_{0}^{\infty} mc_i c_k f I^\alpha \, dI \,
        d\mathbf{c}= \rho v_i v_k + (p+\Pi) \delta_{ik}, \\
& F_{llk} = \int_{R^{3}} \int_{0}^{\infty} mc^2 c_k f I^\alpha \, dI \,
        d\mathbf{c} = \left(5(p+\Pi)+ \rho v^2\right)v_k, \\
        & G_{llk}= \int_{R^{3}} \int_{0}^{\infty} m \left(c^2+\frac{2I}{m}\right) c_k f I^\alpha \, dI \,
        d\mathbf{c} = (\rho v^2 + 2\rho \varepsilon +2p +2 \Pi)v_k,  \\
        & P_{ll}  =\hat{P} _{ll}=\int_{R^{3}}
            \int_{0}^{\infty} m C^2
            Q(f) \, I^\alpha \, dI \, d\mathbf{C}.
            \end{split}
\end{align}

From the balance equations of momentum and of energy in continuum  mechanics, we know that
\[
F_{ik}=\rho v_i v_k -t_{ik}, \qquad  G_{llk}= (\rho v^2 + 2 \rho \varepsilon) v_k - 2t_{ik} v_i+2q_k,
\]
where 
$t_{ik}= -p \delta_{ik} + \sigma_{ik}$ is the stress tensor, $\sigma_{ik}= -\Pi \delta_{ik}+ \sigma_{<ik>} $ is the viscous stress tensor,   $\Pi$ is the dynamical pressure (non equilibrium pressure), $\sigma_{<ik>}$ is the shear stress tensor that is a deviatoric tensor (traceless) and $q_k$ is the heat flux.
Comparing with the closure (\ref{chiuso})$_{(1,3)}$,
 we conclude that the closure gives a result that in the 6-moment theory $\sigma_{<ik>}=0$ and $q_k=0$.  This is the  expected result that there exist no shear viscosity and no heat conductivity in the $6$-moment theory.
 For what concerns (\ref{chiuso})$_2$, taking into account the Galilean invariance with $c_i=C_i+v_i$, we obtain the zero-velocity of $F_{llk}$: 
 \[
 \hat{F}_{llk} = \int_{R^{3}} \int_{0}^{\infty} mC^2 C_k f I^\alpha \, dI \,
         d\mathbf{C} =0.
 \]
Concerning the production term (\ref{chiuso})$_4$, 
the main problem is that, in order to have explicit expression of the production, we need a model for the  collision term, which is, in general, not easy to obtain in the case of polyatomic gases.
With (\ref{chiuso}) we obtain the differential system of $6$ moments:
\begin{align}
\begin{split}
& \frac{\partial \rho}{\partial t}
 +\frac{\partial}{\partial  x_i}( \rho v_i ) = 0, \\
& \frac{\partial \rho v_j}{\partial t} + \frac{\partial }{\partial x_i}\left[(p + \Pi)\delta_{ij} + \rho v_i v_j\right] = 0, \label{eq:ET6}\\ & \frac{\partial}{\partial t} ( 2\rho \varepsilon +\rho v^2 ) + \frac{\partial}{\partial x_i}\left\{ \left[2(p+\Pi) + 2\rho \varepsilon + \rho v^2 \right]v_i \right\} = 0, \\
& \frac{\partial}{\partial t}\left[ 3(p+\Pi) +\rho v^2 \right] + \frac{\partial}{\partial x_i}\left\{ \left[5(p+\Pi) + \rho v^2 \right]v_i \right\}
=\hat{P}_{ll}.  
\end{split}
\end{align}
This system with the thermal and caloric equations of state  (\ref{thermal}) is a closed system for the 6 unknowns ($\rho,v_i,T,\Pi$), provided that we know the collision term in (\ref{chiuso})$_4$.
These results are in perfect agreement with the results derived from the phenomenological theory \cite{NonLinear6}, where a possible expression of the production term $\hat{P_{ll}}$ is adopted. In the case of the BGK approach we obtain $\hat{P_{ll}} =- 3 \Pi/\tau$.

\sz11
\setcounter{chapter}{3}                           
\BSs{}{3.3 Entropy density}

Let us study the entropy density $h$ with non-linear distribution function:
\begin{equation*}
 h = -k_B \int \int f \log f I^\alpha \mathrm{d}I\, \mathrm{d}\boldsymbol{C} 
 = \frac{k_B}{m}\rho \left(\frac{D}{2}- \log \Omega\right), 
\end{equation*}
with $\Omega$ given by (\ref{xizeta})$_3$.
The equilibrium part of the entropy density $h_E$ is expressed as
\begin{align}
 &h_E =\rho s =  \frac{k_B}{m}\rho \left(\frac{D}{2} - \log \Omega_E\right).
\end{align}
Moreover we may notice that the chemical potential $g=\varepsilon+ \frac{p}{\rho} - T s$ is expressed as
\begin{align}
 &\frac{g}{T} = \frac{k_B}{m}\left(1+ \log \Omega_E\right).
\end{align}

On the other hand, the non-equilibrium part of the entropy  is expressed as
\begin{align}
  k= \frac{1}{\rho}(h-h_E) = -\frac{k_B}{m} \log \frac{\Omega}{\Omega_E}.
\end{align}
Since
\begin{align}
 &\frac{\Omega}{\Omega_E} = \left(1+\pop \right)^{-\frac{3}{2}}\left(1 - \frac{3}{D-3}\pop\right)^{-\frac{D-3}{2}},
\end{align}
$k$ is expressed as 
\begin{align}
 &k = \frac{k_B}{2m} \log\left[\left(1+\pop\right)^{3}\left(1 - \frac{3}{D-3}\pop\right)^{D-3}\right].
\end{align}
This expression also coincides with the one
 obtained by the phenomenological ET approach \cite{NonLinear6}.
 $k$ depends on a single variable $Z=\Pi/p$; for $D>3$, $k$ exists and is bounded in the domain that contains the equilibrium state:
 \begin{equation} \label{domain}
 -p <\Pi < p \frac{D-3}{3},
 \end{equation}
 in which $k(Z) <0, \forall\,  Z \neq 0$ and $k$ has a global maximum $k(0)=0$ at the equilibrium state.
 Therefore the convexity condition is satisfied and the ratio between the dynamical pressure and the equilibrium one $\Pi/p$ satisfy the inequalities  \eqref{domain}. According to the Theorem proved in \cite{Serre}, the entropy $h$ has, as is expected, the maximum value at the equilibrium state where $h= \rho s$.

\sz11
\setcounter{chapter}{3}                           
\BSs{}{3.4 Main Field and Symmetric System}
Taking into account (\ref{xizeta}), (\ref{defval}) and (\ref{lambdalino}), we obtain the full expression of the Lagrange multipliers after some cumbersome calculations:
\begin{align}\label{main6}
 \begin{split}
 &\lambda = - \frac{g}{T} - \frac{k_B}{m}\log \frac{\Omega}{\Omega_E}+\frac{v^2}{2T}\left(1+\frac{\Pi}{p}\right)^{-1},\\
 &\lambda_i= -\frac{v_i}{T}\left(1+\frac{\Pi}{p}\right)^{-1}, \\
 &\mu_{ll} =  \frac{1}{2T}\left(1 - \frac{3}{D-3}\pop\right)^{-1},\\
 &\lambda_{ll} = - \frac{1}{2T}\frac{D}{D-3}\frac{\Pi}{p}\left(1+\pop\right)^{-1}\left(1-\frac{3}{D-3}\pop\right)^{-1}.
 \end{split}
\end{align}

According to the general theory developed in \cite{Boillat-1997} for monatomic gases and in \cite{Annals} for polyatomic gases, the components of the Lagrange multipliers coincide with the components of the \emph{main field} \cite{RugStrumia} for which the original system (\ref{eq:ET6}) becomes symmetric hyperbolic. Notice that, in equilibrium where $\Pi=0$, the first five components of the main field  (\ref{main6}) coincide with those obtained by Godunov for the  Euler fluid \cite{godunov}:
\[
\lambda|_E = -\frac{1}{T}\left( g-\frac{v^2}{2}\right), \quad \lambda_i|_E=-\frac{v_i}{T}, \quad \mu_{ll}|_E=\frac{1}{2T},
\]
while $\lambda_{ll}|_E =0$ according to the fact that the Euler fluid is a \emph{principal subsystem} of the 6-moment system \cite{Arma}.
In the reference \cite{NonLinear6}, we proved that the system (\ref{eq:ET6}) can be equivalent to the one obtained many years ago by   Meixner  \cite{Meixner1,Meixner2} via the internal-variable procedure.
In the same paper \cite{NonLinear6}, it was also proved that, in the limit of monatomic gas $D\rightarrow 3$, the system (\ref{eq:ET6}) converges to the Euler system provided that initial data are compatible with the case of monatomic gases, i.e., we should choose $\Pi(0,\mathbf{x})=0$.

We observe that, if we apply the so-called Maxwellian iteration \cite{Ikenberry}, the last equation of (\ref{eq:ET6})  with the production given by the BGK approximation reduces to the Navier-Stokes equation in the absence of the shear stress \cite{6fields,6fieldsMeccanica,NonLinear6}:
\begin{equation}\label{miter}
\Pi=- \nu \,  \text{div} {\mathbf{v}}, \quad \text{with} \quad  \nu=\frac{2}{3}\frac{D-3}{D}p\tau,
\end{equation}
where $\nu$ is the bulk viscosity. The system (\ref{eq:ET6}) in which   the last equation is replaced by (\ref{miter}) was studied by Secchi \cite{Secchi} and by Frid and Shelukhin \cite{Russo,    Hermano}.
Shock wave structure was recently studied for this model  by Taniguchi, Arima, Ruggeri and Sugiyama \cite{ET6shock} showing an excellent agreement with experimental data.

\sz11
\setcounter{chapter}{3}                           
\BSs{}{3.5 K-condition, acceleration waves, and global smooth solutions}
We now want to prove that the so-called Shizuta-Kawashima K-condition \cite{Kawa} is satisfied by our differential system and therefore, according to the general theorems \cite{HN,Wong,Serre}, in contrast to Euler fluid, there exist global smooth solutions provided that initial data are sufficiently smooth.
The system (\ref{eq:ET6}) is a particular case of a generic system of balance laws:
\begin{equation} 
\frac{\partial\mathbf{F}^0(\mathbf{u})}{\partial t}+\frac{\partial\mathbf{F}%
^{i}(\mathbf{u})}{\partial x^{i}}=\mathbf{f(u)}.  \label{sh}%
\end{equation}

We recall that the system (\ref{sh}) satisfies the    K-condition  if, in 
 the equilibrium manifold,   any right characteristic eigenvectors ${\bf d}$ of (\ref{sh})
are not in the null space of $\nabla {\bf f}$, where $\nabla\equiv \partial/\partial {\bf u}$:
\begin{equation}\label{Kcond}
\left.\nabla {\bf f}\,\, {\bf d}\right|_E \neq 0 \quad \forall {\bf d}.
\end{equation}
Lou and Ruggeri \cite{jie} noticed a connection between the K-condition and the global existence of acceleration waves and they propose a necessary weaker K-condition  requiring  (\ref{Kcond}) only for the right eigenvectors corresponding to genuine non linear waves.

For a  quasi-linear hyperbolic system, it is possible to consider a particular class of solutions that characterizes the so-called
\emph{weak discontinuity waves} or, in the language of continuum mechanics, \emph{%
acceleration waves}.
Let us study a moving surface (wave front) $\Gamma$ prescribed by the Cartesian equation $\phi(x,t)=0$ that separates the space into two
subspaces. Ahead of the wave front we have a known unperturbed field $%
\mathbf{u}_0(x,t)$, and behind an unknown perturbed field $\mathbf{u}(x,t)$. Both the fields $\mathbf{u}_0$ and $\mathbf{u}$ are supposed to be 
regular solutions of (\ref{sh}) and to be continuous across the surface $%
\Gamma$, but to be discontinuous in the normal derivative, i.e.,
\begin{equation}
[\mathbf{u}] = 0,   \quad \left[\displaystyle{\frac{\partial \mathbf{u}}{\partial{\phi}}}\right] = {\bm \Pi} \neq 0,  \label{ilj}
\end{equation}
where the square brackets indicate the jump at the wave front.
In \cite{jie} it was verified  that the K-condition is equivalent to the relation: 
\begin{eqnarray*}
\delta\mathbf{f}|_E = \left(\nabla \mathbf{f}\cdot \delta \mathbf{u}%
\right)|_E \propto \left(\nabla \mathbf{f} \cdot \mathbf{d}\right)|_E \neq 0, 
\end{eqnarray*}
where  the operator $\delta$ is defined by $ \delta= \left[{\partial}/{\partial\phi}\right]$.
By introducing the material time derivative, 
the system \eqref {eq:ET6} in the BGK approximation can be rewritten as 
\begin{align}
 \begin{split}
  &\dot \rho + \rho\frac{\partial  v_k}{\partial  x_k} = 0,\\[3pt]
  &\rho \dot v_i+\frac{\partial  }{\partial  x_i}(p+\Pi)
  = 0,\\[3pt]
  &\rho  \dot \varepsilon
  +(p + \Pi) \frac{\partial  v_k}{\partial  x_k}=0, \\[3pt]
    &\left(\frac{p+\Pi}{\rho}-\frac{2}{3}\varepsilon\right)^{\bullet}= -\frac{\Pi}{\rho \tau}.
 \end{split}\label{field_eq_gene0}
\end{align}
As is well known, the characteristic velocities $U$ and the right eigenvectors can be obtained from the system (\ref{sh}) by utilizing the chain rule of the operators:
\[
\frac{\partial}{\partial t} \rightarrow -U \delta, \qquad \frac{\partial}{\partial x_i} \rightarrow n_i \delta, \quad \mathbf{f} \rightarrow 0,
\]
and, in particular,
\[
\bullet \rightarrow - V \delta, \quad V= U-v_n, \quad v_n = v_i n_i.
\]
In the present case, from the  system (\ref{field_eq_gene0}), we obtain
\begin{align} \label{333}
\begin{split}
& -V \delta \rho + \rho \delta v_n = 0, \\
& -\rho V \delta \mathbf{v} +\mathbf{n}\delta(p+\Pi) =0, \\
& -\rho V \delta \varepsilon+(p+\Pi)\delta v_n=0,  \\
& -V \delta \left(\frac{p+\Pi}{\rho}-\frac{2}{3}\varepsilon\right)=0.
\end{split}
\end{align}
Taking into account the constitutive equations (\ref{thermal}) and evaluating (\ref{333}) in an equilibrium state, we have
\begin{equation}
\text{1)}\quad \qquad \,\, V=0 \qquad \longleftrightarrow \qquad  U=v_n, \quad \text{contact waves},
\end{equation}
with $\delta \rho$, $\delta \mathbf{v}_T$, $\delta p$ arbitrary (multiplicity $4$) and $\delta v_n=0$, $\delta \Pi= -\delta p$ ($\mathbf{v}_T$ denotes the tangential velocity);
\begin{equation}\label{sound}
 \text{2)} \,\, V=\pm \sqrt{\frac{5}{3}\frac{p}{\rho}} \quad \longleftrightarrow \quad U=v_n \pm \sqrt{\frac{5}{3}\frac{p}{\rho}}, \quad \text{sound waves},
\end{equation}
with $\delta \rho$ arbitrary, 
\[
\delta \mathbf{v}= {\mathbf{n}V \frac{\delta\rho}{\rho}}, \quad 
\delta \varepsilon =\frac{2}{D}\frac{\varepsilon}{\rho}\delta\rho, \quad 
\delta{\Pi}=\frac{4}{3 D^2}(D-3)\varepsilon\delta \rho.
\]
We notice that the sound velocity in (\ref{sound}) is independent of the degree of freedom $D$ and coincide with the sound velocity of monatomic gas. This curious fact was explained by a general theorem \cite{Annals} in which was proved that  for particular choice of $(N,M)$ systems in which belong the $6$ moment theory the characteristic velocities are independent on $D$.
 
As only the last component of the production term $\mathbf{f}$ of the generic system (\ref{sh}) is non-zero (see (\ref{field_eq_gene0})), the K-condition (\ref{Kcond}) is satisfied if $\delta \Pi \neq  0$.  This is true for contact wave and for sound waves. Therefore the K-condition is satisfied and, together with the convexity of the entropy, we can conclude that, according to the general theorems, the 6-moment system has global smooth solutions for all time and the solution converges to the equilibrium one provided that the initial data are sufficiently smooth.

\setcounter{chapter}{4}                           
\setcounter{equation}{0} 
\BSs{4. Conclusions}{}
In the present paper, we deduced the system of equations for a dissipative fluid in which the dissipation is due only to the dynamical pressure. The closure was obtained by the method of the Maximum Entropy Principle without assuming that the  processes are near equilibrium. This system is the simplest example of  non-linear dissipative fluid after the ideal case of Euler. The system is symmetric hyperbolic with the convex entropy density and the K-condition is satisfied. Therefore, in contrast with the Euler case,  there exist global smooth solutions provided that the initial data are sufficiently smooth. The result obtained here is in perfect agreement with the one obtained by using only phenomenological theory of ET \cite{NonLinear6}. The comparison with experimental data in the case of shock waves is excellent \cite{Taniguchi-2013}.

\lhead[\small\thepage\fancyplain{}\leftmark\hfil$[$]{}
\rhead[]{\small \fancyplain{}\rightmark\small\thepage} \cfoot{}
\markboth {\hfill{\small \rm \Bauthor}\hfill} {\hfill {\small \rm
\Bshorttitle} \hfill}
\renewcommand\bibname{\centerline{\large\bf References}}
\fontsize{9}{11.0pt plus1pt minus .8pt}\selectfont

\textbf{Acknowledgments}: This work was  supported   National Group of Mathematical Physics GNFM-INdAM and by University of Bologna: FARB 2012 Project \textit{Extended Thermodynamics of Non-Equilibrium Processes from Macro- to Nano-Scale}
The author thanks Takashi Arima for the interesting discussions on the contents of this paper.

\end{document}